# BLADE: An Automated Framework for Classifying Light Curves from the Center for Near-Earth Object Studies (CNEOS) Fireball Database

Elizabeth A. Silber[1,*], Vedant Sawal[1]

[1]Sandia National Laboratories, Albuquerque, NM, 87123, USA

*Corresponding author: esilber [at] sandia.gov

**Abstract**

Fireballs (bolides) are high-energy luminous phenomena produced when meteoroids and small asteroids enter Earth's atmosphere at hypersonic speeds, often resulting in fragmentation or complete disintegration accompanied by significant energy release. The resulting bolide light curves capture temporal brightness variations as these objects traverse increasingly dense atmospheric layers, providing essential information on meteoroid entry dynamics, fragmentation behavior, and atmospheric energy deposition processes. The Center for Near-Earth Object Studies' (CNEOS) continuously expanding fireball database offers a globally comprehensive archive of bolide events, including light curves and associated metadata. Events associated with infrasound detections allow direct correlations between acoustic signatures and light-curve features, therefore enabling detailed analyses of fragmentation dynamics and energy deposition. Here, we introduce BLADE (Bolide Light-curve Analysis and Discrimination Explorer), a robust and high-fidelity framework specifically designed to analyze bolide light curves for objects detected from space. BLADE incorporates a processing pipeline integrating Savitzky–Golay filtering, prominence-based peak detection, and gradient analysis, enabling systematic identification and classification of fragmentation events and their associated energy release characteristics. Preliminary results demonstrate that BLADE reliably distinguishes distinct bolide behaviors, providing an objective, scalable methodology for characterization and analysis of large bolide light curve datasets. This foundational work establishes a novel pathway for advanced bolide research, with promising applications in planetary defense and global atmospheric monitoring. Future research should adopt an integrative approach combining CNEOS optical data with complementary infrasound measurements, further clarifying relationships between bolide energy deposition and acoustic signatures, thus refining our understanding of meteoroid and asteroid atmospheric entry processes.

## 1. Introduction

Bright fireballs and bolides are high-energy luminous phenomena produced when meteoroids or asteroids enter Earth's atmosphere at hypersonic speeds (Ceplecha et al., 1998; Silber et al., 2018). The high energy collisional interactions of these extraterrestrial objects with progressively denser atmospheric layers result in intense heating, ablation, and often fragmentation (Bronshten, 1983; Jenniskens, 2004; Popova et al., 2000). Here, we use the terms fireball and bolide interchangeably. Fireballs provide a unique observational window into processes such as atmospheric energy deposition, fragmentation dynamics based on compositional and structural properties, and shock wave morphology and propagation (e.g., Borovička et al., 2017; Gritsevich and Koschny, 2011; Jenniskens et al., 2009; Peña-Asensio et al., 2024; Pilger et al., 2020; ReVelle, 1976; Silber et al., 2018). Moreover, fireballs offer critical clues about the flux rates and internal structure of Near Earth Asteroids (NEAs), significantly advancing risk assessment and mitigation strategies, a subject of considerable scientific and practical importance (e.g., Bailey et al., 2025; Boslough et al., 2015; Boslough and Crawford, 2008; Brown et al., 2013; Chow and Brown, 2025; Harris and Chodas, 2021; Robertson et al., 2024; Sanchez et al., 2024). Beyond their evident scientific relevance, understanding bolide dynamics has practical implications, from planetary defense to interpreting global-scale atmospheric observations (e.g., Landis and Johnson, 2019; Le Pichon et al., 2013; Mainzer, 2017; Pilger et al., 2015; Silber et al., 2011; Wheeler et al., 2024).

Among the many observables associated with these events are the light curves, represented as the time series of a bolide's brightness (Hoffleit, 1933). These photometric profiles are particularly valuable as they encode information about entry dynamics, energy deposition, and fragmentation behavior, as well as the altitudes over which these processes occur (e.g., Koten et al., 2006; Nemtchinov et al., 1994; Pecina and Koten, 2009). Energy deposition is a critical parameter that not only defines the intensity of atmospheric interactions but also influences the generation and propagation of infrasound signals (ReVelle, 1976). Infrasound, a low-frequency acoustic wave generated during high-energy atmospheric events, can travel thousands of kilometers (Christie and Campus, 2010; Evans et al., 1972; ReVelle, 1976). As low frequency acoustic signals provide information related to a bolide's energy release and fragmentation dynamics, infrasound can serve both as an independent observational modality and as a complementary dataset to optical light curves (e.g., Borovička et al., 2013; Silber et al., 2015; Wilson et al., 2025). Accurate interpretation of light curves is therefore essential for reconstructing the dynamical process involved in atmospheric entry, linking energy deposition to acoustic signatures, and understanding the variability in bolide behaviors across altitudes (e.g., Brown et al., 2011; Brown et al., 2013; Silber, 2024b; Silber et al., 2025; Wilson et al., 2025).

Energetic bolides exhibit significant variability in entry velocities, sizes, and compositions, leading to diverse atmospheric interactions and energy deposition patterns (e.g., Borovička et al., 2020a; Brown et al., 2016; Devillepoix et al., 2019; Peña-Asensio et al., 2022; Silber, 2024b; Silber et al., 2025).

Understanding the modes of energy deposition at particular altitudes is vital for interpreting infrasound signals, which are often used to monitor bolides and other high-energy atmospheric events at a global scale (e.g., Mas-Sanz et al., 2020; Ott et al., 2019; Pilger et al., 2020; Wilson et al., 2025). These signals carry information about the dynamics of energy release but require contextual information, such as that provided by light curve measurements, for more comprehensive interpretation. This study is motivated by the need to improve rapid discrimination and characterization of the large volume of natural and artificial events that generate light emissions at high altitudes. For example, artificial objects such as space debris reentering Earth's atmosphere can exhibit fragmentation and energy deposition signatures analogous to those produced by natural objects, necessitating a robust framework for distinguishing between the two populations (e.g., Bowman et al., 2025; Clemente et al., 2025; Ishihara et al., 2012; Silber et al., 2024; Yamamoto et al., 2011).

Previous studies have focused primarily on smaller meteor samples observed by all-sky camera networks over specific geographic regions, often targeting meteor showers or localized populations (e.g., Beech, 2007; Beech and Murray, 2003; Koten et al., 2006; Spurný and Koten, 2007). While these works have advanced our understanding of meteoroid dynamics, particularly regarding light curves, their scope has inherent geographical and dataset size constraints, limiting the generalization of findings to the broader and more diverse range of globally observed bolide events. In contrast, the Center for Near-Earth Object Studies (CNEOS) fireball database[1], maintained by National Aeronautics and Space Administration's (NASA's) Jet Propulsion Laboratory (JPL), offers an unparalleled opportunity global-scale bolide investigations. Aggregating decades of observations from U.S. Government (USG) sensors, the database provides comprehensive metadata on bolide events, including light curves, geographic locations, altitudes, and energies. Unlike localized datasets, the continuously expanding CNEOS database captures the full diversity of energetic bolides, making it particularly well-suited for investigating energy deposition modes, altitudinal distributions, and potential correlations between light curves and infrasound signatures (e.g., Borovička et al., 2020a; Brown et al., 2016; Ens et al., 2012; Gi and Brown, 2017; Silber, 2024b; Silber et al., 2025).

This study is motivated by the need for improved characterization of bolides and their energy deposition, processes critical to understanding fragmentation and energy transfer during atmospheric entry. Using the publicly available light curves hosted on the CNEOS website[2], we develop and introduce BLADE, *Bolide Light-curve Analysis and Discrimination Explorer*, a framework designed to automate light-curve classification, systematically determine the modes and altitudes of energy deposition. By leveraging the CNEOS dataset, this work aims to improve understanding of bolide fragmentation dynamics and, as future work, their relationship to infrasound signals. BLADE

---

analyzes each brightness–time record and rapidly identifies statistically significant fragmentation signatures. Because the workflow relies exclusively on the recorded signal, it scales readily from individual events to the entire catalog, generating a uniform, reproducible classification framework, which can readily be adapted to other high-cadence optical datasets such as those provided by the Geostationary Lightning Mapper (GLM)[3] (e.g., Jenniskens et al., 2018; Ozerov et al., 2024). The study employs an innovative numerical approach utilizing Savitzky-Golay filtering for noise reduction (Savitzky and Golay, 1964). Although the Savitzky-Golay filter has seen extensive use in fields such as spectroscopy (e.g., Chen et al., 2014; Mouazen et al., 2006; Zimmermann and Kohler, 2013), medicine (e.g., Acharya et al., 2016; Hargittai, 2005), geophysics (e.g., Liu et al., 2016), and astronomy (e.g., Li and Paczyński, 2006), its application to bolide light curve classification represents a novel contribution. Because BLADE treats each event in isolation and performs only deterministic operations, identical inputs invariably yield identical outputs, ensuring reproducibility across the entire archive. The resulting feature set forms an immediately exploitable bridge between optical observations and physics-based models of shock formation, infrasound propagation, and post-entry meteorite recovery.

This paper is organized as follows: Section 2 provides an overview of the CNEOS database and its data products and outlines the methodology and numerical approach developed for light curve analysis. Section 3 presents the results obtained from the analysis, demonstrating the capabilities of the proposed approach. In Section 4, we discuss these results. Finally, Section 5 offers conclusions and recommendations for future work.

## 2. Methods

Here, we describe BLADE, an integrated framework comprising data preprocessing, Savitzky-Golay filtering for noise reduction, peak detection, gradient analysis, and event classification. To implement BLADE, we rely on robust, globally distributed data records that capture the wide range of bolide events necessary for accurate classification and analysis.

### 2.1 CNEOS Fireball Database

The CNEOS fireball database is widely regarded as one of the most comprehensive observational archives available for studying meteoroid entry events and their interaction with Earth's atmosphere. Managed by NASA's JPL, the database aggregates observations from USG sensors, which detect bolides on a global scale through their radiative emissions in optical and infrared wavelengths (Nemtchinov et al., 1997). Due to its global coverage, high temporal resolution, and robust energy estimates, the CNEOS database plays a vital role in both planetary defense and global monitoring (e.g., Borovička et al., 2020a; Mainzer, 2017; Mas-Sanz et al., 2020; Peña-Asensio et al., 2022; Silber, 2024b).

---

[3] GLM Bolide Observations: https://neo-bolide.ndc.nasa.gov/#/

The CNEOS fireball database records parameters for each detected fireball, forming the foundation for quantitative and comparative analyses. The event time is reported in Coordinated Universal Time (UTC), enabling temporal correlation between bolide events and complementary datasets, such as infrasound and seismic detections. Accurate timing is essential for determining the propagation delays of acoustic waves and for reconstructing meteoroid trajectories (e.g., Silber, 2024a; Wilson et al., 2025). The geographic location (latitude and longitude) specifies the approximate point of peak brightness, which often corresponds to a fragmentation event or airburst. The altitude, reported in kilometers above sea level, represents the height at which the meteoroid or asteroid reaches peak brightness. This parameter is critical for modeling energy deposition and understanding the generation of shock waves and acoustic energy. Some events also include the pre-atmospheric velocity vectors (with x,y,z components), enabling reconstruction of meteoroid trajectories and estimation of their origin within the solar system or beyond (e.g., Brown et al., 2016; Brown and Borovička, 2023; Devillepoix et al., 2019; Hajduková et al., 2024; Peña-Asensio et al., 2022). It is worth noting, however, that none of the CNEOS fireball quantities (e.g., impact velocity and location), currently include uncertainty estimates; therefore, these quantities and any derived parameters should be treated with appropriate caution in quantitative interpretations. Nevertheless, as a fully signal-driven classification framework, BLADE's performance does not critically depend on the accuracy of these ancillary parameters.

Two energy-related parameters are included in the database. The total radiated energy, derived from optical and infrared emissions, reflects the energy released as visible light during the fireball's luminous phase (Nemtchinov et al., 1997). This radiated energy serves as a proxy for the overall energy of the event and is scaled to estimate the impact energy, expressed in kilotons of trinitrotoluene (TNT) equivalent (1 kt TNT = $4.184 \cdot 10^{12}$ J). The impact energy, calculated using empirically established relationships (Brown et al., 2002), represents the total energy released during atmospheric entry, including the meteoroid's kinetic energy. The spatial and temporal coverage of the CNEOS database is global, encompassing several decades of observations. This extensive temporal coverage facilitates identification of patterns in bolide occurrence, energy distributions, and spatial clustering. However, the resolution of specific parameters, such as altitude and geographic coordinates, depends on sensor geometry and detection conditions; thus, event data uncertainties may exist, particularly in sparsely monitored regions.

Currently, the CNEOS fireball database contains approximately 850 light curves documenting bolide events detected by USG. These light curves are distributed in PDF format, requiring digitization prior to analysis. While future releases of the database may include fully digitized light curves, the currently available records include essential metadata such as the event date and time, location (when available), intensity in counts as a function of time, conversion factors to physical units (W/sr), and total radiated energy in joules. For this study, we selected a subset of events that included coincident infrasound detections, prioritizing bolides observed through multiple modalities to enable future integrative studies.

The initial subset included 144 events, each having infrasound detections published in the literature (Ens et al., 2012; Gi and Brown, 2017; Hupe et al., 2024; Ott et al., 2021; Ott et al., 2019; Pilger et al., 2020; Silber et al., 2011). After careful evaluation of data quality and completeness in both optical and acoustic records, the final subset totaled 124 bolides (Figure 1) (also see Data Availability section). The light curve data were obtained directly from the CNEOS database[2] and digitized[4] for analysis. Preprocessing steps included converting intensity counts to physical units and standardizing the data format to ensure analytical consistency, thus preparing the dataset for subsequent numerical analysis.

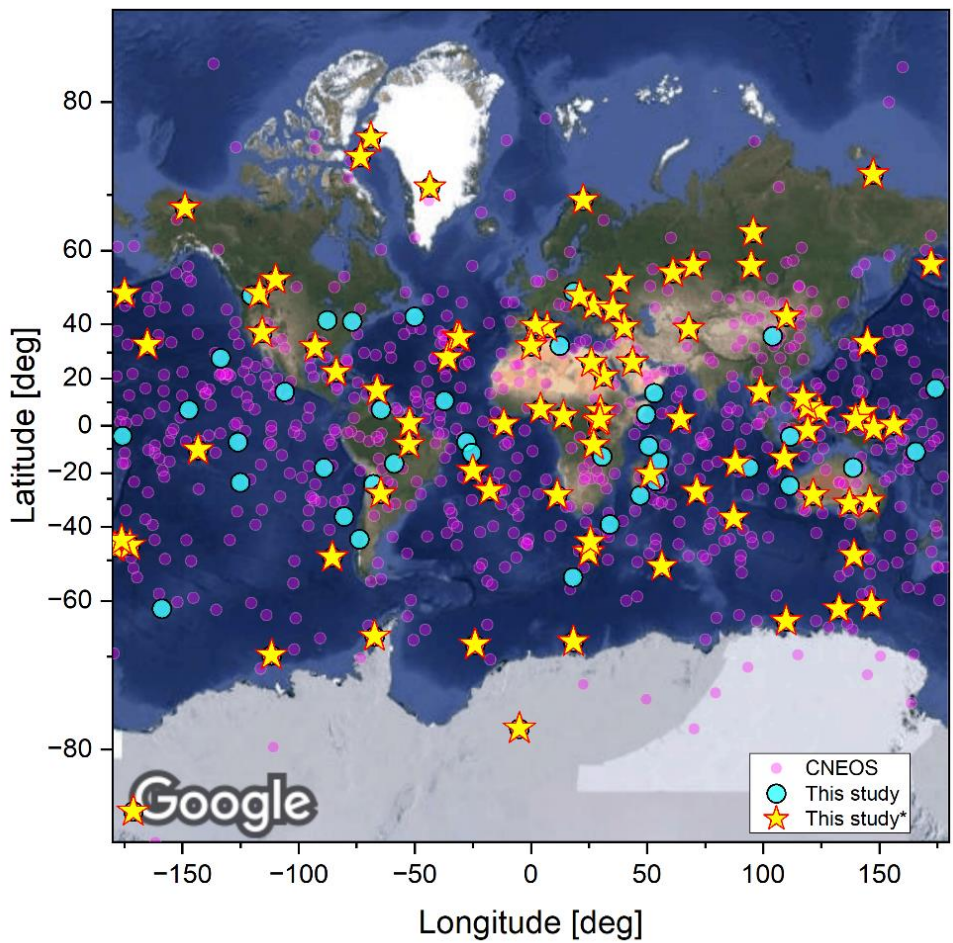

Figure 1. Global detections of bolides reported on the CNEOS website from 1988 to 2024 (plotted are 808 events, which have geographical location). Cyan circles and yellow stars show all events within the subset selected for this study, with the latter having the velocity vector available.

---

[4] Plots were digitized using web-based open source platform (version 4): https://apps.automeris.io/wpd4

## 2.2. Light Curve Classification Algorithm

The classification of light curves in BLADE was conducted in two stages: an initial qualitative analysis followed by a subsequent automated numerical framework. First, an initial visual inspection of bolide light curves was undertaken to preliminarily identify characteristic modes of energy deposition, including continuous fragmentation, discrete fragmentation, and airbursts. This manual analysis served as an initial reference for evaluating and validating the numerical classification algorithm. Given the relatively modest dataset available for this study, the current analysis represents an exploratory effort, laying the groundwork for future rigorous and scalable investigations.

### 2.2.1 Data Preprocessing

As part of BLADE's preprocessing pipeline, light curve data were smoothed and normalized to ensure robust and consistent results. The Savitzky-Golay filter (Savitzky and Golay, 1964), widely used for its effectiveness in reducing noise while preserving significant features such as peaks, was applied to the intensity data.

To quantify the high-frequency noise in each bolide light curve, we compute the standard deviation of its discrete first differences (Oppenheim and Schafer, 2009; Press et al., 2007). Let the raw intensity time series be

$$I = [I_1, I_2, \ldots, I_N] \qquad (1),$$

sampled at a uniform cadence $\Delta t$. We form the first-difference series

$$\Delta I_i = I_{i+1} - I_i, \ i = 1, \ldots, N-1 \qquad (2),$$

which captures the point-to-point fluctuations in the signal. The noise level, $\sigma_{\Delta I}$, is then estimated as the population standard deviation of the $\Delta I$ series:

$$\sigma_{\Delta I} = \sqrt{\frac{1}{N-1} \sum_{i=1}^{N-1} (\Delta I_i - \overline{\Delta I})^2} \ , \text{ with } \overline{\Delta I} = \frac{1}{N-1} \sum_{i=1}^{N-1} \Delta I_i \qquad (3).$$

Because $\Delta I_i$ has the same units as the original intensities (intensity-units per sample), the resulting $\sigma_{\Delta I}$ likewise retains those units. In our pipeline, the raw intensities are subsequently normalized to the interval [0,1]; hence, $\sigma_{\Delta I}$ becomes a dimensionless measure of relative noise amplitude.

This estimate serves two purposes:

(i) Filter parameter selection: window length and polynomial order of the Savitzky–Golay filter are selected to reliably attenuate fluctuations on the order of $\sigma_{\Delta I}$, while preserving broader, physically meaningful peaks.

(ii) Adaptive thresholding: $\sigma_{\Delta I}$ informs adaptive peak detection thresholds (e.g., setting a minimum prominence proportional to $\sigma_{\Delta I}$), accommodating variations in signal-to-noise ratios (SNRs) across events.

By explicitly characterizing the noise floor in this way, our workflow ensures robust smoothing across a heterogeneous set of light curves without manual re-tuning. In python, the noise can be estimated through "*sigma_noise = np.std(np.diff(intensities))*" (Harris et al., 2020).

To account for varying noise levels in the data, the filter parameters were dynamically adjusted based on an estimated noise level. For high-noise data ($\sigma_{\Delta I} > 1.0$), a shorter window length (15 samples) and polynomial order of 2 were applied. This choice reduced the risk of overfitting noisy fluctuations by simplifying the light curve's structure while still capturing the overall trend in the data. Higher-order polynomials in such cases could inadvertently amplify noise, resulting in false peaks and erroneous interpretations. For moderate-noise data ($0.5 < \sigma_{\Delta I} \leq 1.0$), a window length of 21 samples and polynomial order of 3 were applied. For low-noise data ($\sigma_{\Delta I} \leq 0.5$), a longer window length (31 samples) and polynomial order of 3 were applied. This dynamic adjustment optimizess the balance between noise reduction and the preservation of genuine peak features.

In all cases, the chosen window length is enforced to be odd and to not exceed the length of the light curve (adjusted to $N - 1$, and made odd if necessary). Denoting the raw intensities by $I$ and the smoothed output by $\tilde{I}$, we apply

$$\tilde{I} = SG(I; \text{window\_length}, \text{polyorder}) \qquad (4),$$

where SG is the Savitzky–Golay operator. The adaptively selected parameters ensure that: (i) in high-noise data, noise-driven spikes are not misconstrued as physical features, and (ii) in low-noise data, the filter does not overly broaden genuine fragmentation peaks. After smoothing, each light curve is rescaled to the unit interval [0,1] via

$$I' = \frac{\tilde{I} - \min(\tilde{I})}{\max(\tilde{I}) - \min(\tilde{I})} \qquad (5),$$

preparing the series for subsequent prominence-based peak detection.

This normalization step scales smoothed intensities between 0 and 1, enabling consistent analysis across diverse events. Combined with adaptive filtering, normalization ensures that light curves with varying noise levels are uniformly preprocessed, preserving essential signal characteristics and minimizing distortions. Invalid or missing data points (NaN, Inf) were replaced with zeros to prevent computational errors during processing, ensuring robust handling of incomplete or noisy datasets. Although we did not encounter cases of insufficient data in this study, it is worth noting that for events with sparse data, replacing invalid values with zeros may introduce artifacts, as the absence of real signal can lead to underestimated or skewed results in subsequent steps. Future work should explore more sophisticated gap-filling strategies once higher-cadence data become available, but for the present CNEOS dataset the zero-fill approach provides the most stable and reproducible results. Potential improvements include interpolation of missing values or explicit flagging of problematic cases for targeted manual review.

### 2.2.2 Peak Detection

Significant fragmentation and energy-release events were extracted from each smoothed, normalized light curve $I'(t)$ using SciPy's prominence-based peak detection routine *find_peaks*[5] (Virtanen et al., 2020). In BLADE, this prominence-based approach distinguishes genuine fragmentation events from minor fluctuations. This step is specifically aimed to identify fragmentation events and characterize energy-release dynamics while effectively excluding noise-related artifacts.

The peak-detection method evaluates each local maximum against its surrounding minima to compute a prominence value, defined as the vertical distance between the peak and the highest possible baseline connecting it to a higher peak, ensuring that only features rising conspicuously above local noise levels are selected. Three parameters govern peak detection in our implementation:

(i) Prominence (≥0.10 in normalized units) ensures each candidate peak surpasses its immediate baseline neighborhood by at least 10% of the normalized intensity range. This threshold was empirically determined through a systematic examination of representative fragmentation and airburst profiles.

(ii) Height (≥0.10 normalized units) imposes an absolute floor on peak amplitude, further excluding low-amplitude fluctuations that might otherwise satisfy the prominence criterion alone yet remain near the noise baseline.

(iii) Minimum separation (distance = 5 samples or index units) prevents closely spaced indices from generating multiple detections of a single physical pulse; given the uniform sampling cadence of our data, five points correspond to the shortest resolvable interval between independent fragmentation events.

In practice, both prominence and height thresholds were explicitly set at 0.10 (normalized intensity units), ensuring robust selection of peaks rising at least 10% above the local baseline and exceeding an absolute intensity threshold. Higher thresholds can be employed if more conservative detections are required. A minimum separation of five samples was imposed to prevent multiple detections of a single physical pulse, corresponding, in our uniformly sampled data, to the shortest resolvable interval between distinct fragmentation events. This parameterization provided a robust basis for identifying significant fragmentation and energy-release events essential for subsequent bolide light-curve classification. The selected thresholds represent a careful balance between sensitivity to subtle physical features and robust exclusion of noise-induced artifacts. Future iterations based on larger datasets may further refine these parameters, optimizing detection accuracy across a broader diversity of bolide events.

---

[5] https://docs.scipy.org/doc/scipy/reference/generated/scipy.signal.find_peaks.html

Reproducibility is readily achieved using the following Python function call:

"*peaks, props = find_peaks(I_prime, prominence=0.10, height=0.10, distance=5)*", where I_prime represents the Savitzky–Golay smoothed and normalized light curve (Section 2.2.1). The output "peaks" array gives the sample indices of all detected spikes, and "props" contains the corresponding prominence and height measurements. Although the chosen parameters delivered a reliable balance between sensitivity and noise immunity for the current dataset, users may adjust prominence, height, or distance parameters, such as scaling thresholds relative to noise levels or converting sample separations into time intervals, to accommodate alternative sampling cadences or varying noise environments.

### 2.2.3 Gradient Analysis

Gradient analysis was employed within BLADE to supplement the understanding of bolide dynamics by examining the rate of change in normalized intensity over time (Press et al., 2007). The gradient, defined as the numerical derivative of intensity with respect to time, is given by:

$$G(t) = \frac{dI_{norm}(t)}{dt} \qquad (6),$$

where $G(t)$ represents the gradient at time $t$, and $I_{norm}$ is the normalized intensity. This derivative quantifies how rapidly the intensity evolves, with positive values indicating increasing brightness and negative values denoting declining brightness. Sharp increases in $G(t)$ correspond to rapid intensity variations characteristic of airbursts, whereas more gradual changes typically reflect ongoing fragmentation processes.

In Python (Harris et al., 2020), this computation can be straightforwardly implemented as:

"*gradient = np.gradient(normalized_intensities, times)*", where each element of "*times*" represents observation timestamps (in seconds) and "*normalized_intensities*" is the corresponding intensity scaled to [0,1]. This yields derivatives in units of normalized intensity per second; thus, an empirical threshold of 0.5 corresponds to an intensity increase of half a normalized unit per second.

Alternatively, omitting a time vector, as in "*gradient = np.gradient(normalized_intensities)*" assumes unit spacing between samples, such that each element of the gradient represents the change in normalized intensity per sample interval. Under this convention, the 0.5 threshold denotes an intensity increase of half a normalized unit per sample. For datasets with different cadences or varying noise environments, users may either supply an appropriate time array to recover physical units (normalized-intensity $s^{-1}$) or retain unit-spacing derivatives and adjust the thresholds accordingly.

In the current analysis, a maximum gradient threshold of >0.5 was selected empirically based on sensitivity analyses of representative bolide events within the CNEOS dataset. Specifically, this threshold emerged from systematic tests evaluating its effectiveness in distinguishing genuine fragmentation peaks from background fluctuations. Lower thresholds yielded false detections arising from noise and minor intensity variations, while higher thresholds occasionally missed clearly defined fragmentation events. Thus, the selected gradient threshold of 0.5 represents an optimal and statistically informed compromise, balancing sensitivity (accurate fragmentation identification) against specificity (minimizing false positives). Further refinement of these thresholds and criteria is recommended in future iterations based on larger and more diverse datasets.

### 2.2.4 Event Classification

Leveraging the results of peak detection and gradient analysis within BLADE, events were classified into one of five categories based on peak characteristics and temporal dynamics:

1. Continuous fragmentation: Defined by multiple closely spaced peaks (time intervals <10–15 samples), suggesting sequential fragmentation events. The exact threshold (in sample count) can be adjusted to further refine classification accuracy, particularly in ambiguous cases, as discussed later in this paper).

2. Discrete fragmentation: Characterized by peaks separated by significant time intervals (≥10–15 samples), signifying distinct fragmentation episodes occurring independently.

3. Airburst: A single dominant peak exhibiting a steep rise ($G > 0.5$) and a gradual decline, typically associated with a major explosive disintegration event.

4. Single peak: An isolated peak without substantial adjacent activity, potentially representing either a discrete fragmentation event or a relatively minor airburst.

5. No significant peaks: Light curves without detectable peaks above the noise threshold, though such cases were absent in the dataset analyzed here.

Additionally, borderline events with partially overlapping peaks may require manual review or refined threshold parameters to clarify classification. This categorization scheme effectively captures the diverse fragmentation behaviors observed in bolides, providing inherent flexibility for further refinement and optimization as larger and more varied datasets become available.

### 2.2.5 Altitude Extrapolation

Although not central to classification, BLADE also incorporates an optional altitude extrapolation step for bolides with known velocity ($v$) and velocity components. This enables determination of the entry angle ($\theta$), from which altitudes corresponding to specific light curve features can be derived. By anchoring the altitude calculation at the point of maximum brightness ($h_{peak}$), subsequent altitudes are calculated as a function of time, velocity, and entry angle, following the relation:

$$h(t) = h_{peak} - v(t - t_{peak})sin(\theta) \hspace{1cm} (7).$$

This method assumes constant velocity, which provides a first-order approximation. However, since CNEOS velocity vectors lack published uncertainty estimates (e.g., Brown et al., 2016; Devillepoix et al., 2019), derived altitudes inherently carry an unknown error margin. Previous studies (Peña-Asensio et al., 2022) have identified discrepancies in the reported velocity components from the CNEOS database, suggesting typographical errors as a possible explanation. Independent multi-station analyses using all-sky camera observations have also noted potential velocity discrepancies (e.g., Devillepoix et al., 2019; Hajduková et al., 2024; Peña-Asensio et al., 2024; Spurný et al., 2024), emphasizing that altitudes presented here serve primarily as contextual guides and as demonstrations of BLADE's capabilities rather than accurate trajectory reconstructions. Although specific cases of velocity-component reversals have not been confirmed in the present dataset, the potential for such errors must be acknowledged during interpretation. Despite these uncertainties, the calculated altitudes offer valuable spatial context for understanding light curve features and facilitate analyses of energy deposition and fragmentation dynamics (Silber et al., 2025). Future methodological improvements may incorporate velocity deceleration profiles, allowing for refined altitude estimates in cases where detailed trajectory information is available.

### *2.2.6 Methodological Workflow*

To further illustrate and summarize the BLADE methodological framework described in this section, we provide a workflow diagram (Figure 2). Here, each analytical step is outlined, beginning with initial data ingestion and metadata matching (Step 1), followed by digitization and normalization of the bolide light curves (Step 2). Next, adaptive Savitzky–Golay filtering is applied to reduce noise and preserve important signal features, using parameters dynamically adjusted according to the estimated noise levels (Step 3). Prominence-based peak detection then identifies significant peaks within the smoothed, normalized light curves (Step 4). Gradient analysis subsequently quantifies the rate of intensity change (Step 5), enabling robust classification into predefined fragmentation categories (Step 6).

An optional altitude extrapolation step (Step a), requiring velocity-vector information, can be included to determine the altitude of each identified peak by leveraging the provided altitude at peak brightness as an anchor. Computed altitude data are then integrated after gradient analysis to assign altitude values corresponding to each detected peak (Step b). Primary parameter settings and criteria used at each analytical step are explicitly indicated, facilitating transparency and reproducibility of the entire workflow. This visualization also serves as a practical reference for implementing BLADE, allowing users to systematically replicate the workflow and readily adapt parameters to accommodate alternative datasets or specific analytical needs.

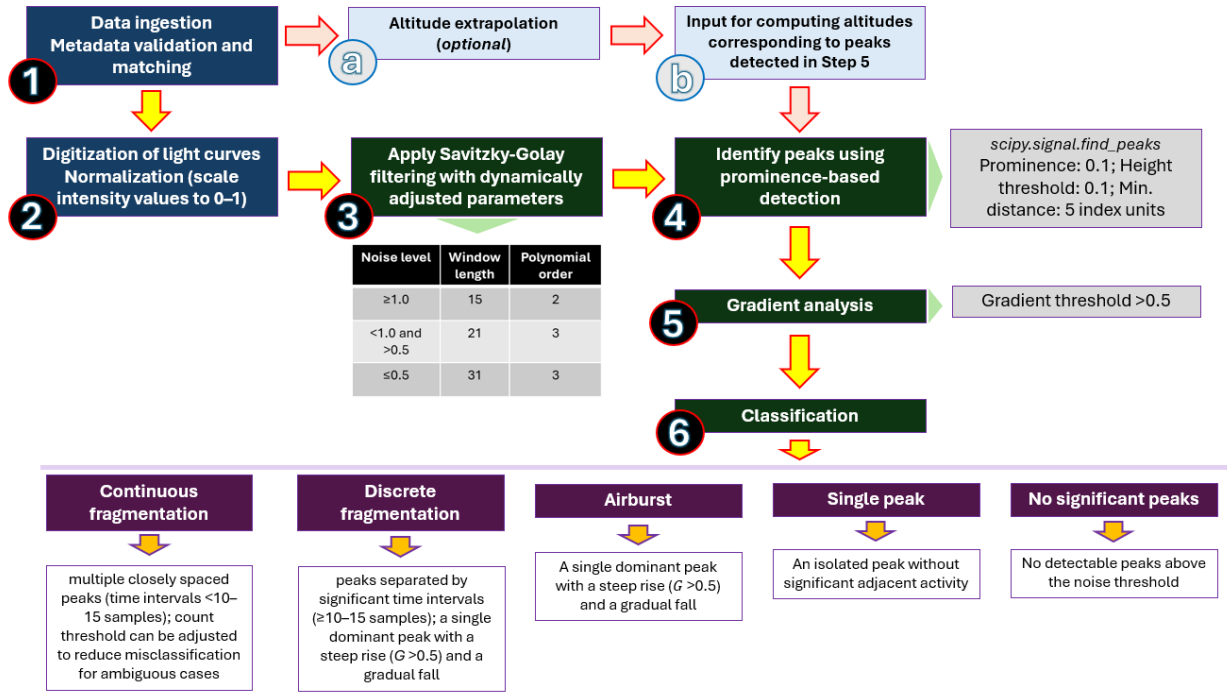

Figure 2: Schematic workflow diagram summarizing the BLADE algorithmic framework. Each analytical step is indicated sequentially, beginning with data ingestion and metadata matching (Step 1) from the CNEOS database, followed by digitization, normalization (Step 2), adaptive Savitzky–Golay filtering (Step 3), prominence-based peak detection (Step 4), gradient analysis (Step 5), and subsequent classification into defined fragmentation categories (Step 6). An optional altitude extrapolation step (Step a), requiring velocity-vector data, can be performed concurrently and then integrated after gradient analysis (Step b) to assign altitudes to detected peaks, using the known altitude at peak brightness as an anchor. Parameter settings employed at each stage, including Savitzky–Golay filtering parameters and peak detection thresholds, are explicitly indicated. The resulting fragmentation classification categories (continuous fragmentation, discrete fragmentation, airburst, single peak, and no significant peaks) and their defining criteria are also shown.

## 3. Results

BLADE demonstrated robust performance in analyzing bolide light curves, including automated peak detection and differentiation, determination of peak timings, and altitude derivation for events with sufficient metadata. The framework successfully classified the analyzed bolide according to the categories defined in Section 2.2.4. For the dataset of 124 light curves, the entire process was completed in a few minutes for the whole set, demonstrating both computational efficiency and scalability.

The preprocessing stage of BLADE effectively reduced noise and enhanced signal features through the application of Savitzky-Golay filtering and normalization. This step successfully preserved important characteristics of the bolide light curves, such as prominent intensity peaks and steep gradients, for subsequent analysis. Prominence-based peak detection reliably identified meaningful peaks, while gradient analysis captured the rate of intensity changes, enabling robust light curve classification. Together, these methods enabled reliable classification of light curves into categories (continuous fragmentation, discrete fragmentation, single peaks, and airbursts). Additionally, the methodology allowed for the determination of altitude–intensity correlations in events where velocity and entry-angle data were available.

Out of 124 analyzed light curves, only nine events exhibited ambiguous classification outcomes between continuous and discrete fragmentation. These ambiguities primarily arose due to variations in the selected sample threshold for peak separation, combined with intrinsic complexity of certain bolide light curves that displayed characteristics consistent with both fragmentation modes. Nevertheless, in all cases, fragmentation was correctly identified as the primary process, demonstrating the robustness of the methodology. This result underlines the capability of BLADE to reliably detect and classify fragmentation processes, even when subtle distinctions in event categorization are present.

Representative examples of bolide events were selected to demonstrate the functionality and effectiveness of the preprocessing and analysis algorithms. These examples encompass a range of energy deposition scenarios, including single airburst events, discrete fragmentation sequences, and altitude-dependent intensity variations. By analyzing these representative cases, we demonstrate BLADE's robust capability to handle diverse and complex light curve morphologies, maintaining consistency in identifying and categorizing relevant features and fragmentation signatures.

### 3.1 Discrete Fragmentation (Case I)

The event recorded on 2006-10-14 at 18:10:49 UTC exhibits distinct characteristics of discrete fragmentation, as identified by the BLADE framework. Figure 3a displays the intensity versus time profile, where two dominant peaks are observed at approximately 0.2 seconds and 0.9 seconds. These peaks are separated by a marked decline in intensity, indicative of sequential fragmentation events along the bolide's trajectory. The abrupt rise and fall of these peaks reflect rapid energy deposition consistent with sudden fragmentation processes. Figure 3b provides the intensity versus altitude distribution, revealing that the primary peaks correspond to altitudes of approximately 47.5 km and 44 km. This altitude distribution suggests that distinct fragmentation events occurred at these heights, resulting in localized energy release. The normalized intensity curve, shown in Figure 3c, emphasizes the relative contributions of the two main peaks by eliminating absolute intensity differences. This representation stresses the dominance of the fragmentation events while allowing secondary features to emerge more clearly. Such normalization aids in distinguishing primary fragmentation events from smaller-scale energy fluctuations that may arise from secondary processes. Figure 3d integrates gradient analysis with peak detection to provide a detailed view of the fragmentation dynamics. The gradient overlay identifies sharp transitions in intensity, corresponding to the fragmentation events, with the detected peaks marked by red dots. The two primary peaks are clearly resolved, while additional smaller peaks suggest the presence of secondary fragmentation or related phenomena. These minor features reflect the complex interplay of aerodynamic and thermal stresses acting on the bolide as it traversed the atmosphere.

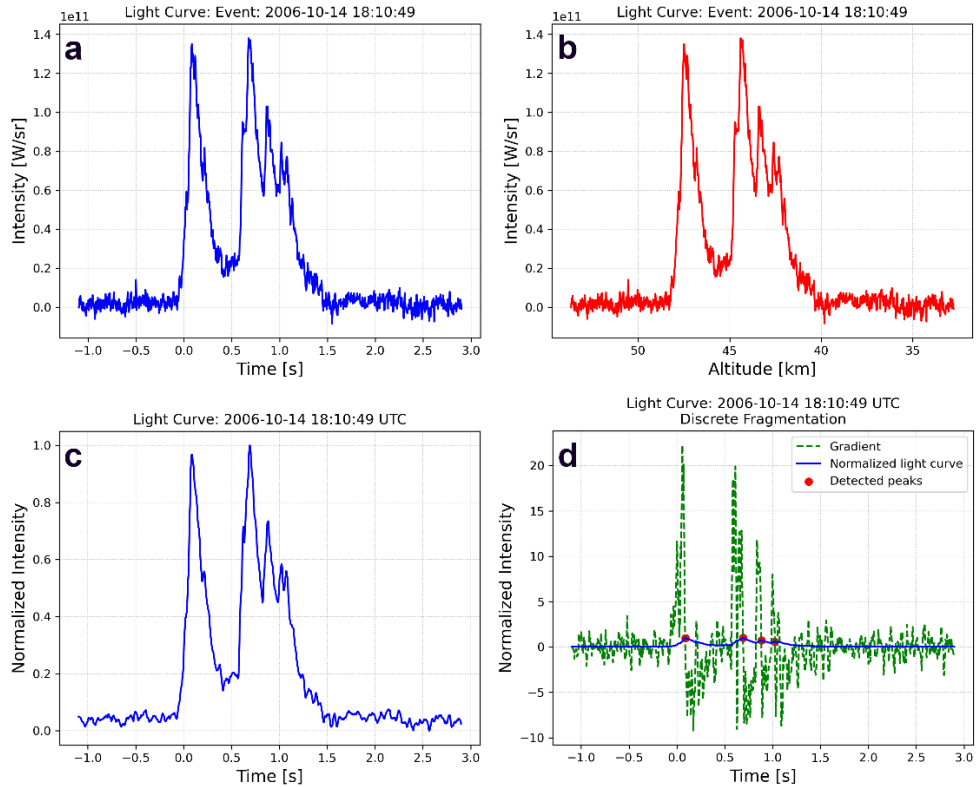

Figure 3: Light curve analysis of the event on 2006-10-14 at 18:10:49 UTC. (a) Intensity versus time plot showing two dominant peaks corresponding to distinct fragmentation events. (b) Intensity versus altitude plot indicating that the primary peaks occurred at altitudes of ~47 km and 44 km, with additional fluctuations suggesting possible secondary fragmentation. (c) Normalized light curve. (d) Gradient analysis and peak detection plot, where the green dashed line represents the gradient, the blue line indicates the normalized light curve, and red dots mark the detected peaks.

### 3.2 Discrete Fragmentation (Case II)

The 2015-06-14 03:03:06 UTC event (Figure 4) presents characteristics indicative of discrete fragmentation, as identified by BLADE. There is a pronounced primary peak at approximately 0.5 seconds, accompanied by smaller secondary peaks on both sides (Figure 4a). This pattern suggests a dominant fragmentation event occurring at the central timepoint, with additional smaller disintegration events contributing to the observed intensity fluctuations. Figure 4b shows the spatial manifestation of this fragmentation process. A primary intensity peak is observed near 32 km altitude, correlating with the most significant energy release. Secondary peaks are evident at approximately 40 km and 28 km altitudes, indicating additional fragmentation events occurring at different heights during the object's descent. The normalized light curve (Figure 4c) shows the dominance of the central peak. The pronounced central peak corresponds to the largest energy deposition, consistent with a significant fragmentation event. The gradient analysis of the normalized light curve is shown in Figure 4d. This analysis confirms the discrete nature of the fragmentation, with distinct peaks detected at 0.2, 0.5, and 0.8 seconds. These peaks align with the features observed in Figures 4a and 4c, validating the algorithm's ability to resolve discrete fragmentation events.

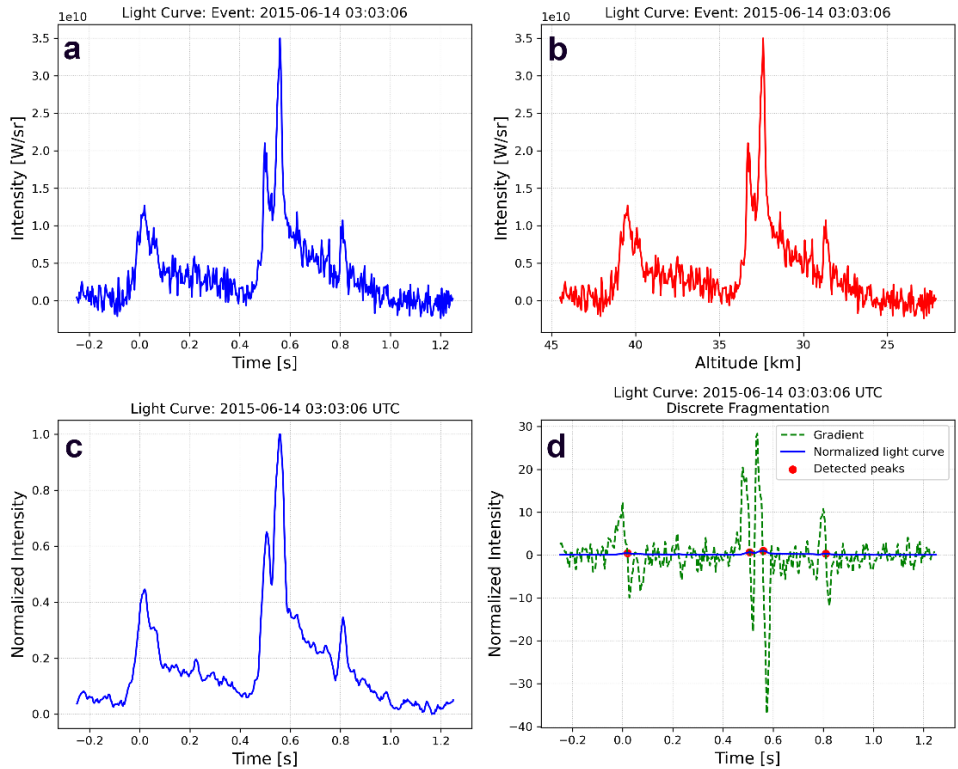

Figure 4: Light curve analysis of the 2015-06-14 03:03:06 UTC event, classified as discrete fragmentation. (a) Intensity vs. time. (b) Intensity vs. altitude shows the primary fragmentation at ~30 km, with secondary events at ~35 km and ~28 km. (c) Normalized light curve emphasizes the prominence of individual peaks. (d) Gradient and detected peaks confirm discrete fragmentation.

### 3.3 Continuous Fragmentation (Case I)

Figure 5 displays the results for the event that occurred on 2015-05-10 at 07:45:01 UTC. It was classified as 'continuous fragmentation' event by the algorithm. We selected this event because it is also an example of an ambiguous case, exhibiting features typical of both continuous and discrete fragmentation. This ambiguity arises due to overlapping peaks and intensity distributions, which will be further discussed in the next section. The intensity versus time plot (Figure 5a) shows a broad, multi-peaked structure spanning approximately 0.4 seconds. The largest peak occurs around 0.25 seconds, followed by a secondary peak at approximately 0.4 seconds, indicative of overlapping or closely spaced fragmentation events. The intensity versus altitude plot (Figure 5b) reveals a gradual increase in intensity from around 31 km altitude, culminating in a primary peak at approximately 30 km, and a secondary peak near 27 km. These observations suggest a progressive fragmentation process as the bolide descended. In Figure 5c, the normalized light curve emphasizes the relative dominance of the primary fragmentation event at 0.25 seconds while capturing the weaker, subsequent fragmentation signatures. The smoothed curve further clarifies the temporal distribution of energy deposition, corroborating the continuous nature of the fragmentation. Figure 5d shows the gradient analysis and peak detection results. The gradient (green dashed line) captures the intensity variations, while the red markers identify several detected peaks along the normalized light curve (blue line). The bolide exhibited the continuous fragmentation behavior, with energy deposition occurring over an extended duration and across multiple altitudes.

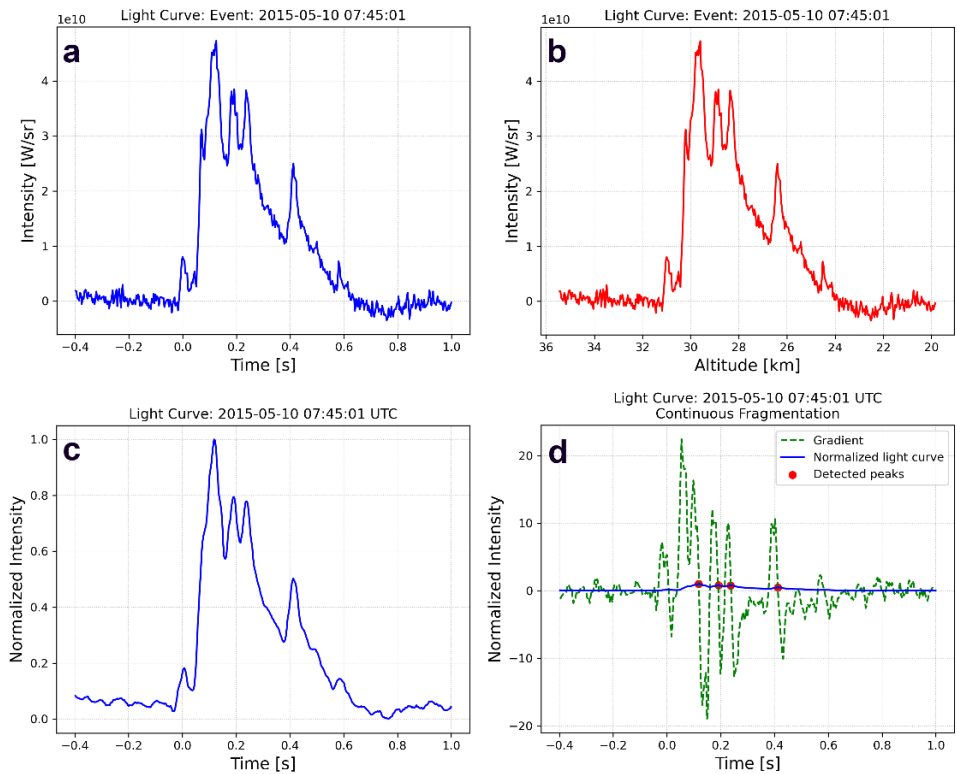

Figure 5: Analysis of the 2015-05-10 07:45:01 UTC event demonstrating continuous fragmentation. (a) Intensity vs. time plot showing a multi-peaked structure. (b) Intensity vs. altitude plot highlighting primary and secondary peaks at ~30 km and ~27 km, respectively. (c) Normalized intensity vs. time plot emphasizing relative peak dominance. (d) Gradient analysis with detected peaks (red markers) identifying multiple fragmentation events.

### 3.4 Continuous Fragmentation (Case II)

The light curve analysis for the event on 2019-11-28 at 20:30:53 UTC as identified by BLADE demonstrates a continuous fragmentation process characterized by a sustained energy deposition profile. Figure 6a presents the intensity versus time plot, showing a relatively broad primary peak around 0.2 - 0.4 seconds. The intensity profile is irregular, indicating complex fragmentation dynamics with multiple sub-peaks, suggesting a sequence of smaller fragmentations superimposed on the broader energy release. The profile's asymmetry, with a gradual rise and a more pronounced decay, indicates non-uniform fragmentation processes. Figure 6b displays the intensity versus altitude plot, illustrating the spatial distribution of the energy deposition. The peak intensity is observed near an altitude of approximately 34 km, with secondary features extending between 26 km and 38 km. This altitude range corresponds to the region of maximum atmospheric interaction, where the meteoroid likely experienced continuous fragmentation under increasing dynamic pressure. Figure 6c shows the normalized intensity versus time, offering a clearer representation of the temporal structure. Additional fluctuations corresponding to smaller fragmentation events are discernible throughout the profile. Figure 6d presents the normalized light curve with its gradient overlaid and detected peaks identified. The detected peaks, marked with red dots, correspond to significant intensity variations, showing discrete moments of increased energy deposition within the continuous fragmentation regime. The gradient reveals the rate of change of intensity, providing clues about the dynamics of fragmentation. The observed features are consistent with the expectations for continuous fragmentation, where the meteoroid undergoes progressive disintegration, releasing energy over an extended duration and altitude range.

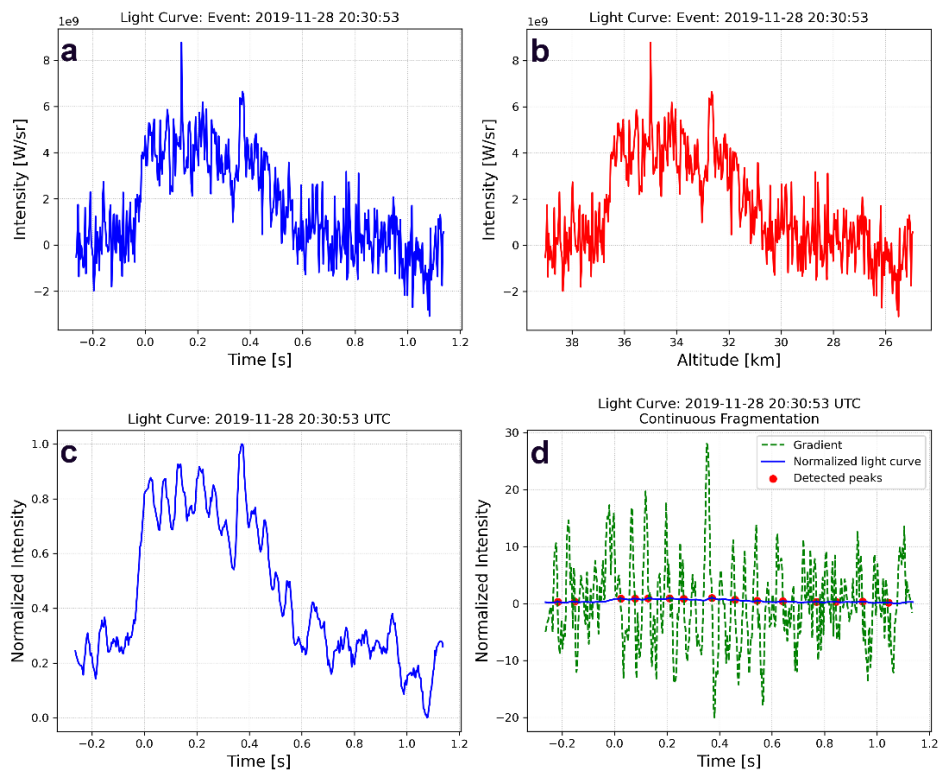

Figure 6: Analysis of the 2019-11-28 (20:30:53 UTC) event, classified as a continuous fragmentation. (a) Intensity versus time plot showing a broad primary peak and multiple sub-peaks, indicative of complex energy deposition. (b) Intensity versus altitude. (c) Normalized intensity versus time plot. (d) Normalized light curve with gradient overlay and detected peaks, showing significant intensity changes and fragmentation dynamics.

### 3.5 Airburst

The analysis of the 2018-12-18 23:48:18 UTC event (Figure 7) illustrates a clear demonstration of an airburst event characterized by a predominant energy release. The light curve exhibits a pronounced peak at approximately 0.35 seconds (Figure 7a). The abrupt rise in intensity, followed by a relatively smooth decline, is indicative of a rapid and concentrated energy deposition, consistent with the physical processes typical of airburst phenomena. This temporal profile suggests the absence of prolonged fragmentation or multiple bursts, as typically observed in other fragmentation cases. The peak intensity occurs at an altitude of ~26 km (Figure 7b), with intensity decreasing symmetrically above and below this altitude. This pattern indicates that the energy was predominantly released at a single point or within a narrow altitude range, aligned with the atmospheric disruption typically observed in airbursts. The sharp gradient in intensity further corroborates the inference of a high-energy detonation or explosion in the atmosphere rather than a gradual disintegration of the bolide. The distinct and singular nature of the light curve seen in Figure 7c illustrates the dominance of a single peak and the absence of additional features that would otherwise indicate discrete or continuous fragmentation. A gradient analysis of the light curve, including the detection of the peak, is shown in Figure 7d. The gradient curve captures the rapid increase in intensity leading up to the peak, as well as the sharp decline afterward.

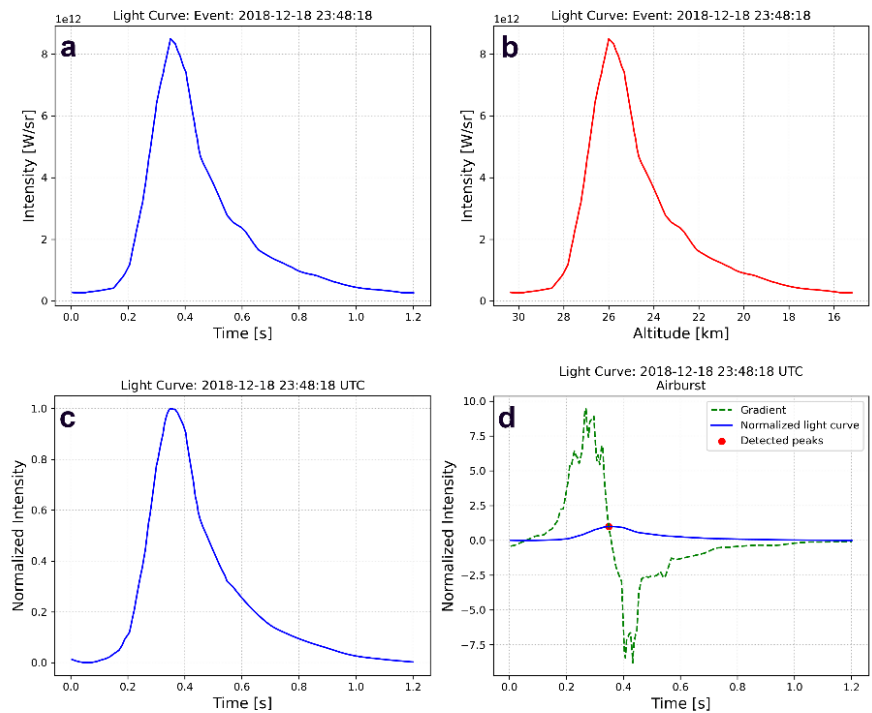

Figure 7: Light curve analysis of the 2018-12-18 23:48:18 UTC airburst event. (a) Intensity vs. time showing a dominant peak at ~0.35 s. (b) Intensity vs. altitude indicating a peak energy release at ~26 km. (c) Normalized light curve. (d) Gradient analysis with the detected peak marked at 0.35 s, confirming a single energy release characteristic of an airburst.

### *3.6 Single Peak*

The light curve analysis for the event on 2022-02-07 (20:06:25 UTC) reveals a single, prominent peak (Figure 8), which was classified as a single-peak event by BLADE. Figure 8a presents the intensity as a function of time, showing a rapid increase to a well-defined peak followed by a symmetric decrease, indicating a singular energy release. The temporal profile lacks secondary fluctuations, reinforcing the single-peak classification. Figure 8b shows that the energy deposition was localized at a specific altitude. This sharp energy release pattern suggests a singular fragmentation or airburst event during the bolide's atmospheric entry. Figure 8c shows the normalized light curve. In Figure 8d, the gradient analysis and detected peak are shown. The sharp gradient changes and the precise temporal detection of the peak are consistent with a dominant, isolated process without evidence of fragmentation within the analyzed time frame. However, while this event is classified as a single-peak light curve, the overall evidence suggests it is better characterized as an airburst. We will revisit this distinction in the next section.

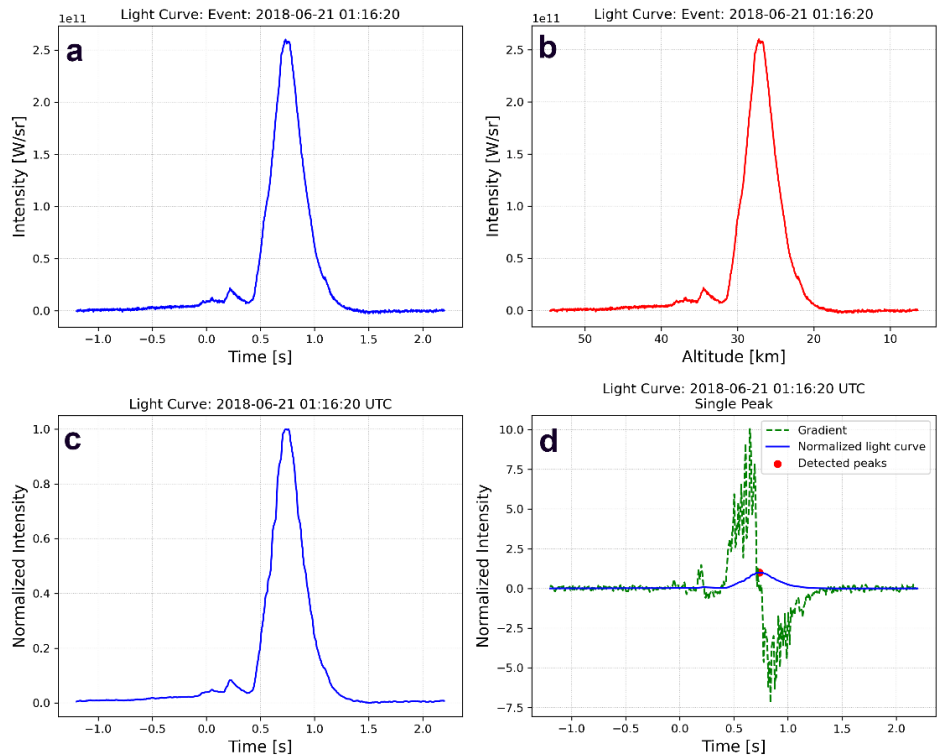

Figure 8: Light curve analysis for the 2018-06-21 01:16:20 UTC event. (a) Intensity vs. time showing a single peak. (b) Intensity vs. altitude indicating localized energy deposition. (c) Normalized light curve. (d) Gradient analysis with detected peak, supporting single-peak classification, though evidence suggests an airburst event requiring further algorithm refinement.

**3.7 Aggregated Results**

While not pertinent to BLADE, we plotted the distribution of bolide events classified by the algorithm into four categories: airburst, continuous fragmentation, discrete fragmentation, and single peak (Figure 9). This distribution primarily serves to illustrate the composition of the analyzed dataset rather than to explicitly evaluate algorithmic performance. The pie chart indicates that the majority of events (65.9%) were classified as discrete fragmentation, indicating that this mode of energy deposition is the most commonly observed in the dataset. Single peak events account for 24.6% of the classifications, airbursts constitute 7.1%, and continuous fragmentation account for only 2.4%, making this category the least frequent in the analyzed sample. This distribution reflects BLADE's effectiveness in systematically categorizing diverse bolide signatures. The relatively high proportion of single peak events may overlap with airbursts, given the similarities inherent in their respective light curve profiles.

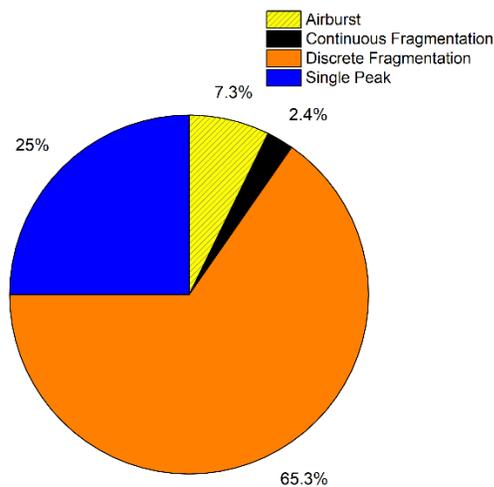

Figure 9: Distribution of bolide events classified into four categories: airburst, continuous fragmentation, discrete fragmentation, and single peak.

Although rigorous comparative analyses of BLADE's classification results are beyond the scope of this initial demonstration, we include selected parameter distributions associated with each classification category to identify preliminary trends and patterns that could motivate future, more detailed investigations. Figure 10 provides a comparative statistical overview of bolide classification categories (airburst, single peak, discrete fragmentation, and continuous fragmentation) across various bolide parameters, including altitude, entry angle, impact energy, and entry velocity. Airburst events typically cluster within a narrow altitude range (~30–40 km), whereas single peak events exhibit a somewhat wider spread (Figure 10a). The altitude distributions of discrete and continuous fragmentation events span broader ranges, reflecting diverse fragmentation dynamics at varying atmospheric layers. Discrete fragmentation events exhibit greater variability in entry angle, indicating varied trajectories and energy dynamics (Figure 10b). Airbursts appear to be more commonly associated with steeper entry trajectories, whereas single-peak events are observed across a broader range of entry angles. Future analyses with an expanded dataset will be necessary to determine whether this trend reflects a genuine physical correlation or arises from classification uncertainties, where some single-peak events may represent airbursts that were not identified by current detection criteria. Both continuous and discrete fragmentation categories tend to occur at somewhat lower impact energies compared to airbursts and single peak events (Figure 10c). Most events cluster around lower velocities (Figure 10d). This clustering likely reflects inherent sampling biases within the USG-detected bolides, predominantly comprising NEAs, and is consistent with previously published average impact velocity estimates for the NEA population (e.g., Brown et al., 2002; Tricarico, 2017).

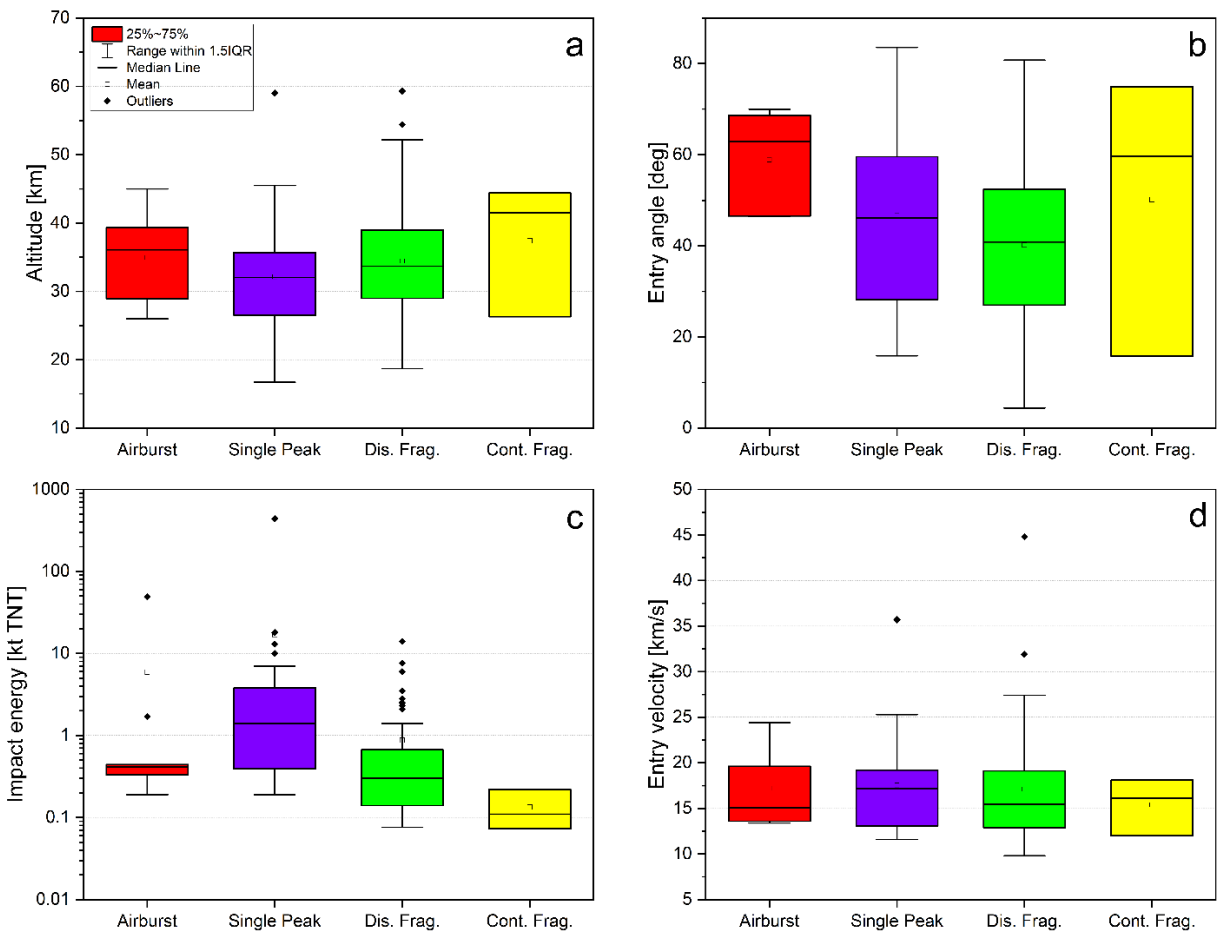

Figure 10: Box-plot statistical comparison of bolide classification categories (airburst, single peak, discrete fragmentation (Dis. Frag.), and continuous fragmentation (Cont. Frag.)) across: (a) altitude, (b) entry angle, (c) impact energy, and (d) entry velocity. Box boundaries represent the lower (Q1) and upper quartiles (Q3), with the central line indicating the median. Whiskers extend to ±1.5 × IQR (interquartile range), and outliers beyond this range are plotted individually. The legend shown in panel (a) is applicable to all panels.

Figure 11 presents an overview of bolide events by exploring relationships between impact energy, altitude, velocity, and entry angle across various classification categories. The distribution of altitude as a function of impact energy shows substantial scattering, although it is evident, as expected, that higher energy events often occur at lower altitudes (Figure 11a). Continuous fragmentation events tend to exhibit lower velocities (Figure 11b). In Figure 11c, discrete fragmentation events span a broader range of velocities and altitudes, reflecting greater variability in their atmospheric interactions. Figure 11d relates the entry angle to impact energy. While discrete fragmentation events appear scattered, airbursts and single peak events generally align with steeper entry angles. These distributions provide context for the classification categories identified in this study, although detailed exploration of these trends lies beyond the current scope.

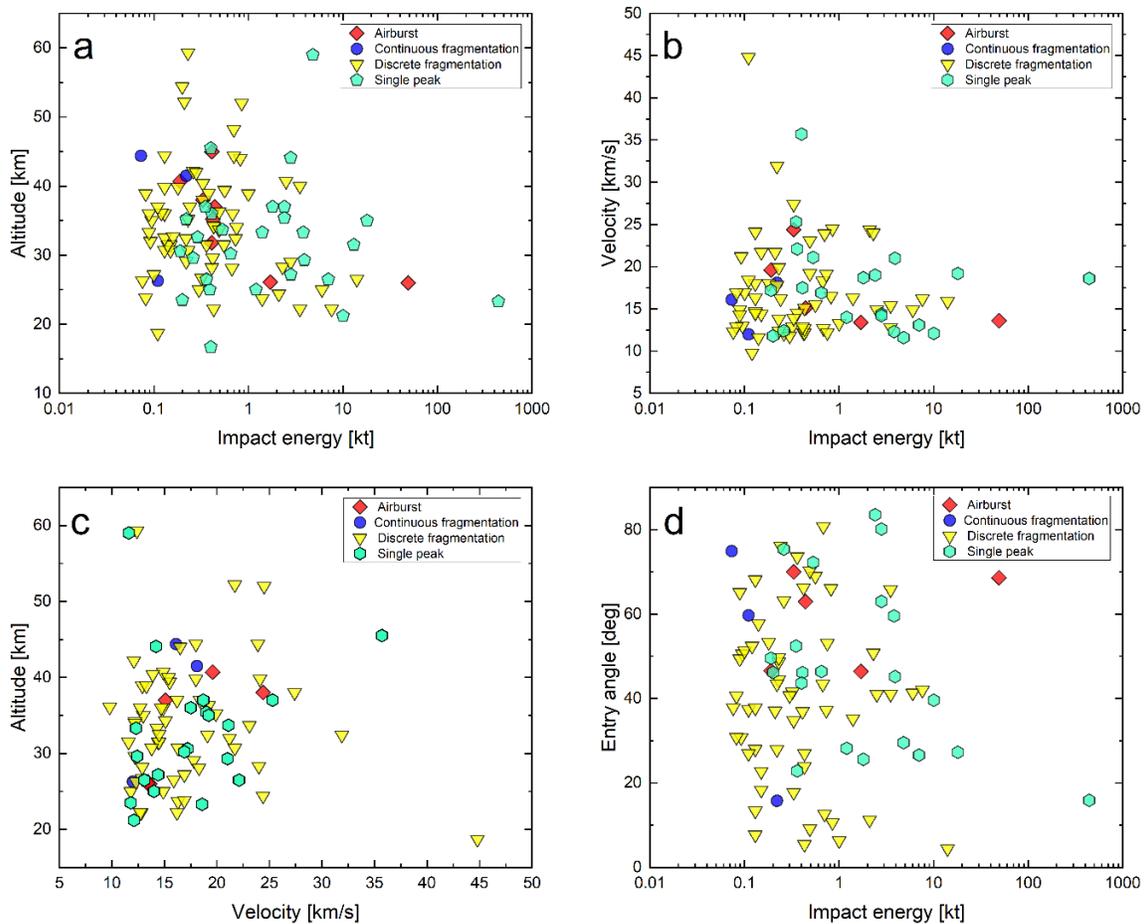

Figure 11: Relationships between impact energy and altitude, velocity, and entry angle across different classification categories.

## 4. Discussion

By leveraging the global and diverse dataset provided by the CNEOS database, this work introduces BLADE (Bolide Light-curve Analysis and Discrimination Explorer) as a robust, scalable framework for systematic light curve analysis, enabling efficient processing of large datasets. Although the present analysis focuses on a subset of light curves generated by a meter-sized and larger class of objects detected from space, the methodologies presented here lay an important foundation for more comprehensive studies. The framework presented in this study demonstrates a novel and systematic approach to analyzing and classifying CNEOS bolide light curves based on their intensity profiles. Furthermore, it provides a basis for differentiating and categorizing luminous signatures generated by artificial hypersonic objects in the upper atmosphere (e.g., space debris). BLADE effectively identifies and distinguishes primary fragmentation processes, such as airbursts, single-peak events, discrete fragmentation, and continuous fragmentation. However, the presence of lower intensity secondary processes, such as gradual disintegration, ablation, or very minor fragmentations, introduces complexities that require further refinement of the classification criteria.

Before further elaborating on the significance of our results, it is useful to briefly contextualize this work relative to prior studies. Previous fragmentation investigations employing semi-analytical modeling approaches (e.g., Borovička et al., 2019; Borovička et al., 2020b; Borovička et al., 2013; Henych et al., 2023) focused extensively on reconstructing entry dynamics, aerodynamic stress profiles, and ablation behavior from detailed event-specific observations. In contrast, BLADE represents a fundamentally distinct methodological approach: it is an automated, data-driven framework specifically developed for objective, systematic classification of bolide fragmentation signatures from extensive observational datasets, such as the CNEOS archive. BLADE does not rely on explicit assumptions regarding aerodynamic parameters, meteoroid structural properties, or ablation mechanisms. While semi-analytical models remain essential for detailed reconstruction of individual events, BLADE's novel contribution is its capacity for large-scale, unbiased fragmentation classification. Integration of BLADE-derived classifications with physics-based entry models, incorporating dynamic pressure and aerodynamic analyses, represents a logical and compelling next step toward comprehensive characterization of fragmentation regimes.

The results indicate that primary processes involving significant energy release predominantly govern the observed bolide light curve profiles. However, secondary features often overlap or modify these primary signatures, complicating their interpretation. For instance, single-peak events, such as the 2022-02-07 case, present ostensibly straightforward intensity profiles yet exhibit subtle signatures consistent with airburst dynamics (Figure 8). Similarly, airbursts, as demonstrated by the 2018-12-18 event, exhibit a dominant singular peak characterized by a steep intensity gradient and a concentrated altitude profile (Figure 7). Although BLADE reliably identifies these as airburst events, secondary fragmentation episodes occurring at lower altitudes may partially influence the observed intensity profile. This emphasizes the need for further refinements in classification criteria to better account for and differentiate such subtle secondary processes.

Fragmentation dynamics, whether discrete or continuous, further illustrate the complexities of bolide atmospheric entry. Discrete fragmentation events, such as the 2006-10-14 and 2015-06-14 cases, were marked by distinct peaks in the intensity profile, each corresponding to a separate fragmentation event (Figures 2 and 3). In contrast, continuous fragmentation, as seen in the 2019-11-28 event, displayed broader intensity profiles with overlapping peaks indicative of sustained disintegration (Figure 6). These distinctions demonstrate the effectiveness of the BLADE framework in classifying fundamental fragmentation dynamics. Nevertheless, continuous fragmentation events often include additional features, such as localized energy releases at varying altitudes, which may resemble secondary airburst-like features. Future refinements involving advanced classification criteria, including more stringent gradient thresholds or spectral analyses, will be essential to further disentangle and accurately characterize these overlapping phenomena.

The analysis also emphasizes the importance of altitude data in understanding the spatial dynamics of energy deposition. BLADE's ability to directly correlate intensity peaks with derived altitudes provides critical information about the vertical distribution of energy deposition across different atmospheric layers. For example, the discrete fragmentation event on 2006-10-14 (Figure 3) exhibited energy release at distinct altitudes (~47 km and 44 km), while the continuous fragmentation event on 2015-05-10 (Figure 5) exhibited a broader altitude range (~34 km to ~27 km). However, such analyses depend on the availability of metadata, such as velocity vectors and peak brightness altitudes, which are currently available for only a fraction of events in the CNEOS database (~30%). Even when these metadata are available, the associated uncertainties are not quantified, limiting the accuracy of derived altitudes and related analyses, as discussed in Section 2.1 (e.g., Brown et al., 2016; Brown and Borovička, 2023; Devillepoix et al., 2019; Hajduková et al., 2024; Peña-Asensio et al., 2022). Expanding the availability and completeness of these metadata will be critical for improving BLADE's capability to derive robust altitude-intensity correlations and refine energy deposition models. Additionally, while the current study focuses on CNEOS bolide light curves, the inherent scalability of BLADE suggests its potential applicability to other high-cadence observational datasets, such as those provided by the Geostationary Lightning Mapper (GLM). Extending BLADE to GLM data would significantly expand the observational capabilities for detecting and analyzing bolide fragmentation events, thereby advancing both temporal coverage and scientific utility.

Classification ambiguities were observed in only nine cases, arising primarily from the intrinsic complexity of certain bolide light curves exhibiting characteristics of both continuous and discrete fragmentation modes. These ambiguities demonstrate the need for a multifaceted classification approach that integrates temporal, spatial, and intensity-based features. Future developments could include advanced machine learning models trained on larger, more varied datasets to improve classification accuracy and effectively handle these complex cases. Events such as the 2015-05-10 (Figure 5) case demonstrate that fragmentation dynamics are not always a binary process of discrete or continuous modes but can exist along a spectrum influenced by factors such as meteoroid composition, entry angle, and atmospheric conditions. For example, a meteoroid with heterogeneous material properties, including the presence of volatiles and high porosity, may simultaneously experience gradual erosion and sudden structural failures, producing light curves with blended features. Similarly, dynamic pressures encountered at lower altitudes can trigger secondary fragmentations or airburst-like phenomena, further complicating the observed light curve profile.

These complexities have significant implications for interpreting energy deposition processes and their spatial distribution. Accurate classification is critical for reliably linking light curve features to infrasound signals, since the mode and altitude of energy deposition directly control the generation and propagation of acoustic waves (e.g., Albert et al., 2023; Le Pichon et al., 2013; ReVelle, 1976; Silber, 2024b). Ambiguous classifications may introduce uncertainties in this correlation, emphasizing the importance of developing more refined classification methods capable of disentangling overlapping features and capturing the full range of fragmentation behaviors. Nevertheless, the robustness and fidelity of the current algorithm are demonstrated by its consistent identification of the dominant fragmentation mode, even when subtle secondary processes occur. In all ambiguous cases, fragmentation was correctly recognized as the primary mechanism of energy deposition, confirming the algorithm's effectiveness in capturing the core dynamical processes. Future improvements of BLADE, potentially incorporating advanced classification criteria or machine learning techniques, could further reduce uncertainties arising from overlapping features and secondary phenomena. Ultimately, these cases emphasize the requirement for increasingly sophisticated classification criteria capable of accurately characterizing complex processes such as gradual disintegration, ongoing ablation, or localized secondary airbursts.

The classification distribution (Figure 9) shows that discrete fragmentation is the most common mode of energy deposition, indicating its predominance among the analyzed bolide events. However, the relatively small proportions of airburst and single peak events suggest potential challenges in distinguishing these categories, particularly given their shared characteristics and limited sample sizes. Expanding the dataset and refining classification criteria will be critical to improving the accuracy of event categorization, thereby providing greater insight into the underlying bolide energy deposition dynamics.

While this study was focused specifically on developing the BLADE framework for bolide light curve characterization and classification, we also explored possible relationships between altitude and impact energy, velocity and impact energy, altitude and velocity, as well as entry angle and velocity (Figure 11). Although no clear, unambiguous trends emerged from the current dataset, future analyses incorporating a wider range of events, including both smaller and significantly larger bolides, may yield more definitive and informative patterns.

Limitations in the dataset stem from the restricted nature of the detection systems that produce the CNEOS bolide light curves. Insufficient transparency regarding the acquisition and processing of these data constrains the detailed interpretation of subtle energy-deposition dynamics captured within the recorded light curves. Moreover, the light curves themselves may not fully reflect the complexity inherent in the physical processes occurring during atmospheric entry. For example, the CNEOS light curve for the Chelyabinsk bolide shows only a single dominant peak, giving no indication of the intricate energy deposition and fragmentation patterns observed in video recordings of the event (Brown et al., 2013). This discrepancy suggests that the light curves in the CNEOS dataset may oversimplify or inadequately capture detailed fragmentation behaviors and energy-release mechanisms. Consequently, integrating additional observational data sources, such as video footage and infrasound measurements, is paramount for achieving a more comprehensive understanding of bolide fragmentation and energy deposition mechanisms.

Finally, the importance of maintaining and expanding the CNEOS bolide database cannot be overstated. Although this study effectively leverages the existing dataset, the absence of post-2022 light curves and limited metadata restricts the potential for further refinement and validation. Facilitating regular and systematic releases of comprehensive bolide light curve data and associated metadata to the broader scientific community will be essential for developing increasingly robust analysis algorithms and advancing our fundamental understanding of bolide entry dynamics.

## 5. Conclusions and Future Work

In this work, we introduced a novel numerical approach for the automated classification of bolide light curves, focusing specifically on events documented in the CNEOS database. By systematically analyzing these data, our methodology provides a robust framework for characterizing energy deposition dynamics and atmospheric entry processes. By integrating Savitzky-Golay filtering, prominence-based peak detection, and gradient analysis, our algorithm effectively classifies bolide light curves primary categories: single-peak events, airbursts, discrete fragmentation, and continuous fragmentation. The results demonstrate the algorithm's ability to reliably extract dominant features from light curves, identify primary processes, and derive corresponding altitudes of energy deposition for events with sufficient metadata.

While the algorithm reliably classifies primary processes, secondary processes, such as overlapping fragmentations, gradual disintegration, or subtle ablation events, introduce complexities requiring further refinement of the classification framework to better capture the nuances inherent in bolide observations. Despite the limitations inherent in the CNEOS fireball database, including incomplete metadata availability and the absence of post-2022 light curves, this study emphasizes the significant potential of the existing dataset as a foundation for advancing bolide research. Expanding access to comprehensive bolide light curve data and associated metadata will be critical for further refining BLADE and supporting broader scientific investigations. With continued algorithmic enhancements and improved data availability, our approach holds considerable promise for significantly advancing understanding of atmospheric entry dynamics, energy deposition mechanisms, and planetary defense strategies.

Future work should therefore prioritize the following directions:

i. Dataset expansion: Incorporating additional light curves with higher temporal resolution and more complete metadata, including velocity vectors.

ii. Algorithm refinement: Developing increasingly sophisticated classification criteria, including spectral analysis, advanced gradient-based thresholds, and the integration of machine learning techniques, to increase classification accuracy and effectively manage complex bolide scenarios.

iii. Validation and benchmarking: Systematically validating and benchmarking the classification algorithm against expanded observational datasets and simulated bolide events to quantify performance and address existing classification ambiguities.

iv. Correlation with infrasound: Extending analysis to integrate bolide-derived infrasound data, enabling comprehensive assessments of energy deposition processes and associated acoustic wave generation.

v. Integration with physics-based entry modeling: Utilizing derived velocity and altitude estimates in conjunction with atmospheric density profiles to calculate dynamic pressure, aerodynamic stress, and inferred meteoroid structural strength. These parameters can then directly inform physics-based modeling approaches to meteoroid and asteroid fragmentation mechanics, ablation processes, and shock wave generation, thus providing a rigorous physical basis for interpreting fragmentation behavior across diverse extraterrestrial object populations.

vi. Application to additional observational platforms: Adapting and expanding the BLADE methodology to accommodate datasets from other high-cadence, space-based optical platforms, such as the Geostationary Lightning Mapper (GLM), thereby extending the observational capabilities and general applicability of the approach.

The continued availability, regular updating, and comprehensive expansion of the CNEOS database will be critical for realizing these objectives. By systematically facilitating open access to bolide light curve data, the scientific community can address existing limitations, optimize classification methodologies, and significantly advance the understanding of atmospheric entry processes. The current study provides a strong foundation for these ongoing efforts, demonstrating the considerable potential of automated light curve analysis to substantially contribute to bolide research and planetary defense objectives.

**Acknowledgements:** We thank the anonymous reviewer for valuable comments that helped improve the manuscript. Sandia National Laboratories is a multi-mission laboratory managed and operated by National Technology and Engineering Solutions of Sandia, LLC (NTESS), a wholly owned subsidiary of Honeywell International Inc., for the U.S. Department of Energy's National Nuclear Security Administration (DOE/NNSA) under contract DE-NA0003525. This written work is authored by an employee of NTESS. The employee, not NTESS, owns the right, title, and interest in and to the written work and is responsible for its contents. Any subjective views or opinions that might be expressed in the written work do not necessarily represent the views of the U.S. Government. The publisher acknowledges that the U.S. Government retains a non-exclusive, paid-up, irrevocable, world-wide license to publish or reproduce the published form of this written work or allow others to do so, for U.S. Government purposes. The DOE will provide public access to results of federally sponsored research in accordance with the DOE Public Access Plan.

**Funding:** This work was supported by the Laboratory Directed Research and Development (LDRD) program (project number 229346) at Sandia National Laboratories, a multimission laboratory managed and operated by National Technology and Engineering Solutions of Sandia, LLC., a wholly owned subsidiary of Honeywell International, Inc., for the U.S. Department of Energy's National Nuclear Security Administration under contract DE-NA0003525.

**Data availability:** The methods section provides comprehensive algorithmic descriptions, parameter definitions, and explicit computational steps, enabling full independent replication of all analyses presented here. While the BLADE source code is not publicly available at this time, it or selected elements may become available in the future. A .csv file containing metadata and output classifications for all bolide events analyzed in this study is included as electronic supplementary material. All original data are freely available via the NASA CNEOS fireball database (https://cneos.jpl.nasa.gov/fireballs/), with raw light curve files specifically downloadable from: https://cneos.jpl.nasa.gov/fireballs/lc/. Light curve digitization was performed using a publicly accessible web-based open-source tool (version 4): https://apps.automeris.io/wpd4/.

**Conflict of interest:** The authors declare no conflict of interest.